# Lean CNNs for mapping electron charge density fields to material properties

*Pranoy Ray[1,2], Kamal Choudhury[3,4], Surya R. Kalidindi*[1,2]*

[1] George W. Woodruff School of Mechanical Engineering, Georgia Institute of Technology, USA
[2] School of Computational Science and Engineering, Georgia Institute of Technology, USA
[3] Materials Science and Engineering Division, National Institute of Standards and Technology, Gaithersburg, USA
[4] Theiss Research, La Jolla, USA

* Correspondence: surya.kalidindi@me.gatech.edu

## ABSTRACT

This work introduces a lean CNN (convolutional neural network) framework, with a drastically reduced number of fittable parameters (<81K) compared to the benchmarks in current literature, to capture the underlying low-computational cost (i.e., surrogate) relationships between the electron charge density (ECD) fields and their associated effective properties. These lean CNNs are made possible by adding a pre-processing step (i.e., a feature engineering step) that involves the computation of the ECD fields' spatial correlations (specifically, 2-point spatial correlations). The viability and benefits of the proposed lean CNN framework are demonstrated by establishing robust structure-property relationships involving the prediction of effective material properties using the feature-engineered ECD fields as the only input. The framework is evaluated on a dataset of crystalline cubic systems consisting of 1410 molecular structures spanning 62 different elemental species and 3 space groups.

## 1. INTRODUCTION

Density Functional Theory (DFT) [1, 2] is often used to estimate molecular systems' physical and chemical properties and facilitates rapid screening of potential chemical compositions for advanced materials design [3, 4, 5, 6, 7]. A DFT simulation utilizes the raw molecular structure of a material as input and self-consistently computes the electron charge density (ECD) field, which is then used to estimate a multitude of the material's effective properties. However, the high cost of the DFT computations often precludes the direct estimation of the material properties over the large spaces of chemical compositions explored in material design efforts.

Emergent AI/ML (artificial intelligence/machine learning) toolsets [8, 9, 10, 11, 12] make it possible to learn the underlying molecular structure-property relationships [13, 14, 15, 16, 17, 18, 19, 20, 21] on a relatively small set of DFT computations and subsequently employ the extracted surrogate models to identify molecular structures [13, 22] of high potential interest (i.e., expected to exhibit a desired combination of material properties) for further analyses by DFT simulations. Indeed, AI/ML tools are already being used extensively in the establishment of Process-Structure-Property (P-S-P) relationships at different material structure length scales [23, 24, 25]. At any selected material structure length scale, especially for mesoscale considerations, it has been shown that the framework of 2-point spatial correlations [26, 27, 28, 29, 30, 31, 32, 33, 34, 35] combined with convolutional neural networks (CNNs [36]) or Gaussian process regression (GPRs [37]) has proven to be extremely valuable in establishing the desired S-P surrogates. However, for continuous local states (i.e., non-eigen microstructures [refs]) such as those studied in the present work, it is well-known that the 2-point correlations do not capture all details of the statistically homogeneous fields. In fact, they only capture the first-order spatial information. Although this feature set is not complete, it nevertheless lies in a convex space, providing valuable guidance on what other values are theoretically possible for the feature set. Recent work has begun to apply these same strategies to molecular structures and has demonstrated promise [15, 22, 36]. One of the main advantages of utilizing spatial correlations is that it provides a systematic quantification framework for the rich defect populations encountered in the multiscale material internal structure. Specifically, the framework allows for the extraction of a hierarchy of feature vectors quantifying the higher-order spatial patterns in the material structure. Furthermore, it can be shown that the complete set of all theoretically possible material structures occupies a convex region in the feature vector space. It is noted that many of the other feature engineering approaches [18, 36, 37, 38, 39] explored in current literature for the material structure, such as graph neural networks [14, 16, 40] or expert-selected feature lists [8, 20, 21, 41, 42], do not guarantee that the physically allowed material structures occupy a contiguous region in the latent vector space. As already noted, convex or contiguous design spaces become important when pursuing materials design efforts, where the primary goal is to evaluate a large number of novel theoretically feasible candidate structures that have not yet been confirmed in practice.

Prior work ([46], [47], [48], [49], [50], [51], [52], [53], [54]) has identified the ECD field (denoted as $\rho$) as the central physics-capturing molecular structure field. One of the main advantages of considering the ECD field as the molecular structure field is that it implicitly embeds elemental species information and exhibits an inherent capability to extrapolate to elemental species that are not present in the training dataset [22, 38]. Past literature has explored the effectiveness of the ECD field as the direct input vector to CNN-based surrogates ([46], [50], [51], [54]). In general, these models needed to employ a large number (up to 10 million) of training parameters, suggesting that these models might represent overfits to relatively small datasets of ~ 2500 samples used in these studies. However, it is expected that effective properties of the molecular structures can be directly related to the spatial correlations extracted from the ECD field. Specifically, it is expected that the formalism of n-point spatial correlations [26] can produce a systematic approach for feature engineering the ECD field. The spatial correlations of a material structure have proven their efficacy for establishing accurate and robust surrogate modeling across multiple material length scales [15, 22, 26, 35, 52, 53, 54], including molecular length scales [22, 38]. Our central hypothesis for this work is that the use of autocorrelations of the ECD field[*] as the input will result in a significantly smaller CNN (a lean CNN without the need for fully connected layers) model compared to those in current literature that utilize the ECD field directly as the input. Similar lean CNNs using autocorrelations of material structures as input features have already demonstrated robust structure-property linkages at the mesoscale [57].

---

[*] Note that the computational cost of obtaining the ECD field from DFT computations can be significantly lower than the cost of estimating the effective properties from the ECD field using the DFT computations. Furthermore, the same ECD field is capable of informing us about multiple effective properties of interest.

The primary goal of this paper is to develop and demonstrate the benefits of building lean CNNs to serve as surrogate models connecting the autocorrelations of the ECD field to the effective properties of interest. We demonstrate these benefits on a small dataset of high structural complexity and chemical diversity. Specifically, our dataset includes 1410 compositions covering 3 space groups, across 62 unique chemical species. The properties targeted in this work include (a) Bulk Modulus and (b) Total Energy of the material system.

## 2. DATA COLLECTION AND PRE-PROCESSING

### 2.1. DATASET

A dataset consisting of ECD fields, bulk moduli, and total energies of approximately 1410 cubic crystalline compounds is compiled from the JARVIS-DFT [58] repository. These properties were calculated using density functional theory (DFT) with the Vienna Ab initio Simulation Package (VASP) ([59], [60], [61]), employing a plane-wave basis set and the projector augmented wave (PAW) method with the OptB88vdW exchange-correlation functional. VASP-recommended pseudopotentials were utilized for each element. The compounds included in this dataset comprised 15 unary, 494 binary, 947 ternary, 93 quaternary, and 1 quinary compound spanning 62 unique chemical species. Figure 1(a) shows the distribution of the compounds in this dataset across three space groups of which $Fm\bar{3}m$ and $F\bar{4}3m$ are FCC (face-centered cubic), and $Pm\bar{3}m$ is SC (simple cubic).

In this study, we selected two very different types of target properties to evaluate critically the versatility of our feature engineering approach (i.e., using autocorrelations of the ECD field as input to our proposed lean CNNs). One of the targets was selected as the bulk modulus (represents an example of a structural property) and the other is total energy (represents a functional property). These properties are physically related to each other by the Birch-Murnaghan equation of state ([62], [63]). Figure 1(b) and (c) represent the distribution of DFT-computed bulk moduli and total energies of the compounds present in the dataset. Note that the samples exhibit an uneven distribution in both the property targets.

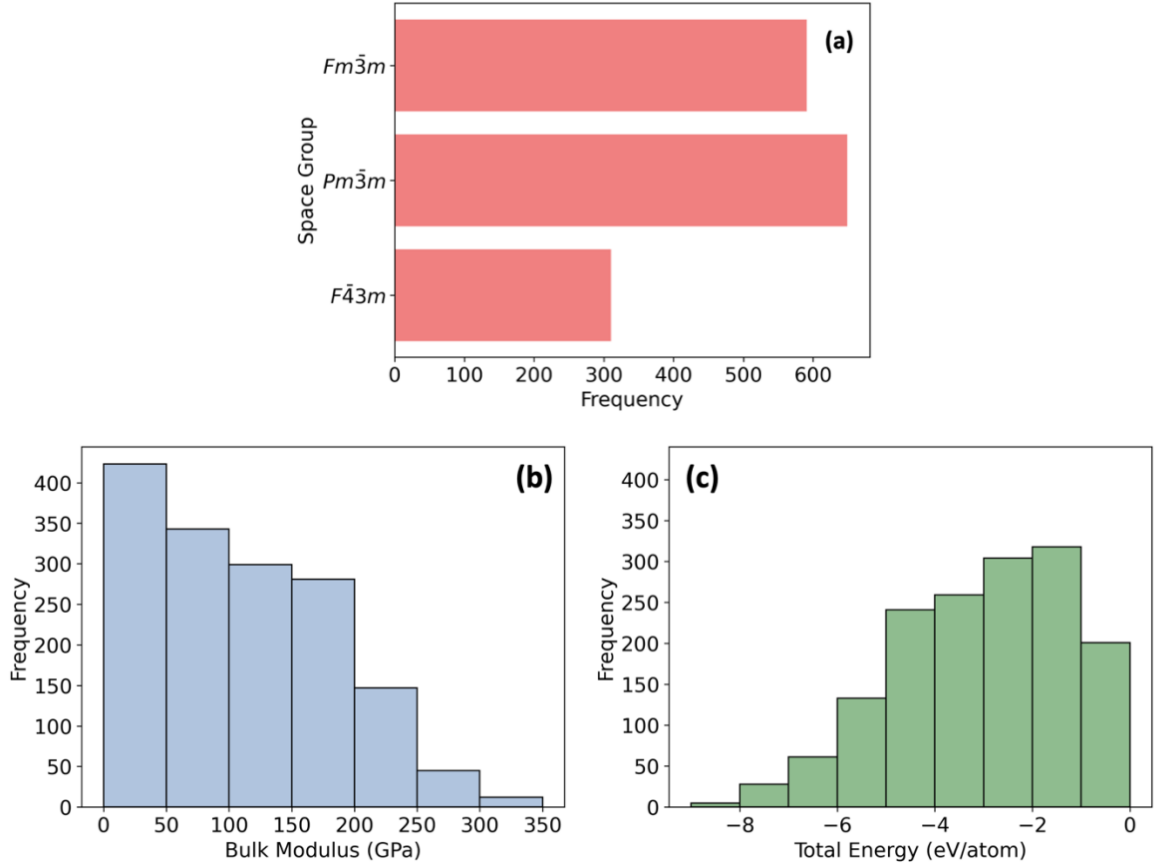

**Figure 1: (a)** Distribution of the three cubic space groups for the different compounds present in the dataset. **(b)** Histogram of the bulk moduli of the compounds in the dataset. **(c)** Histogram of the total energies of the compounds in the dataset.

**2.2. DATA PRE-PROCESSING**

**2.2.1. PRIMITIVE TO CONVENTIONAL CELL TRANSFORMATION**

For reducing the computational cost of DFT simulations, unit cells are frequently depicted using their primitive basis to reduce the number of atoms present while still taking advantage of periodicity. Specifically, the FCC primitive cells in our dataset were primarily represented using a periodic non-orthogonal basis. Consequently, it was necessary to convert the material structure representation from the primitive basis into the conventional (orthogonal & periodic) basis. This conversion was performed using trilinear interpolation, employing the SimpleITK package [64] in Python. The accuracy of this conversion was evaluated by ensuring negligible changes in the peak and minimum values of the ECD and the total number of valence electrons before and after transformation for all compounds included in our study. The SC material systems did not need this conversion as their SC primitive cell was already expressed on an orthogonal basis. The resampled conventional unit cell representation of the ECD (i.e., ECD values sampled on a uniform orthogonal grid) is used as the default ECD from hereon and is denoted by the array $\rho_s$, where the subscript $s$ indexes the location for every voxel in the ECD. Figure 2(a) provides an example of the primitive to conventional ECD field transformation performed on all the primitive FCC ECD fields utilized in this work.

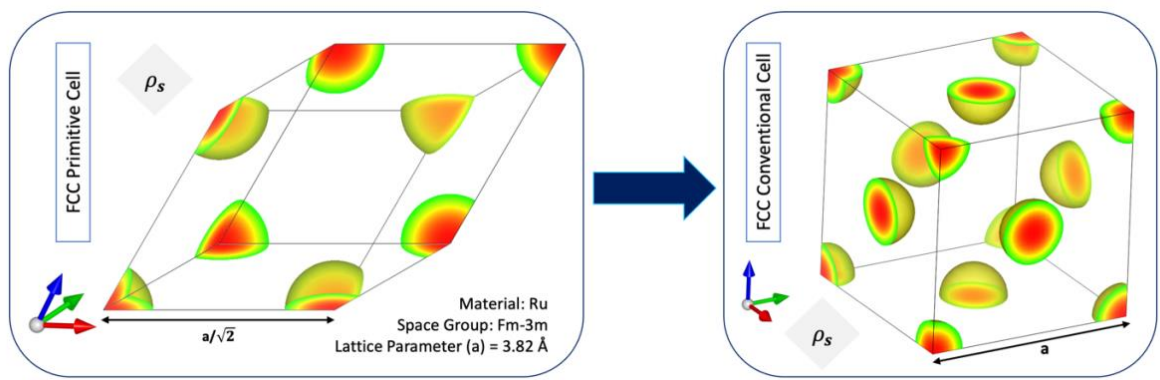

(a) Primitive to Conventional Cell transformation for an example FCC compound.

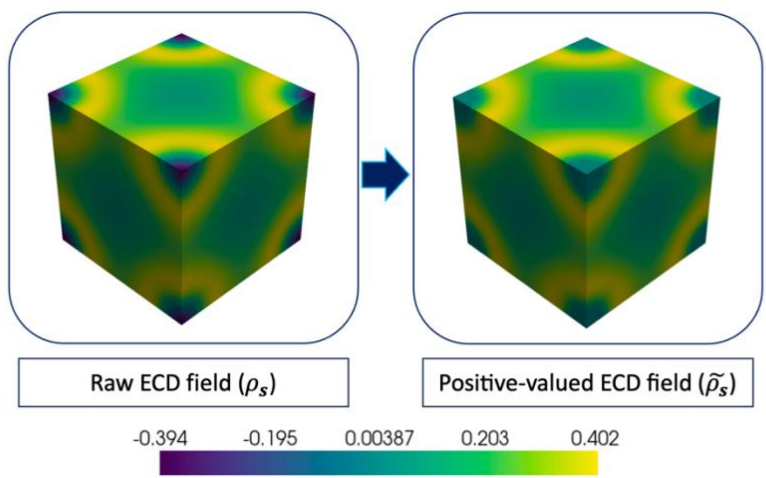

(b) Re-normalization of the ECD Field for an example SC compound.

**Figure 2:** Data pre-processing steps performed on all ECD fields in the dataset.

### 2.2.2. HANDLING NEGATIVE VALUES

In certain ECD fields, we observed regions of negative electronic charge density, especially when using projected augmented wave (PAW) pseudopotentials [65]. This phenomenon occurs because these pseudopotentials take the "soft" charge density—derived from Kohn-Sham wavefunctions—and apply an additional "correction" known as "augmentation" to the charge density. Although this "augmentation" is often described in the literature ([66],[67]) as an addition of charge, it can also result in a decrease in charge density in specific regions. If the soft charge density in a given area is lower than the amount of charge removed by the augmentation, negative charge density values arise. Causes of this effect can be attributed to various factors, but one common cause is Fourier aliasing, which occurs when the augmentation grid is insufficiently refined to accurately represent the augmentation charge density. The ECD fields ($\rho_s$) utilized in this study were sourced from a curated repository where each DFT simulation was run with a different augmentation charge grid. Since negative values for ECD are non-physical, we removed them and re-normalized the ECD values using the protocols described next.

The DFT-computed charge density array, denoted by $\rho_s$ introduced earlier, exhibits the following property:

$$\text{Total Number of valence electrons (in the simulation cell S)} = \sum_{s \in S} \rho_s V_s \quad (1)$$

where $V_s$ denotes the volume of the voxel indexed by $s$ (constant in our case for all voxels in a single DFT computation result). Our goal is to re-normalize the ECD field, i.e., transform $\rho_s$ into positive-valued $\tilde{\rho}_s$, while keeping the same total number of valence electrons in the simulation cell $S$ (i.e., $\sum_{s \in S} \rho_s V_s = \sum_{s \in S} \tilde{\rho}_s V_s$).

We achieve this by first identifying connected regions of negative ECD values in the simulation cell $S$, referred henceforth as clusters ($C$). These clusters typically correspond to individual atoms in the molecule. For each cluster $C_i$, we calculate the sum of the negative ECD values as $\rho_{C_i}^-$. Next, we identify a boundary layer of ECD voxels ($B_j$) immediately outside $C_i$ (the voxels exhibiting positive ECD values). The sum of the positive ECD values in $B_j$ is denoted as $\rho_{B_j}^+$. Where $\rho_{B_j}^+ \leq |\rho_{C_i}^-|$, we zero-out the ECD values in $B_j$, and update $\rho_{C_i}^-$ by adding $\rho_{B_j}^+$ (i.e. $\rho_{C_i}^- = \rho_{C_i}^- + \rho_{B_j}^+$). We repeat this process until we encounter the situation $\rho_{B_j}^+ > |\rho_{C_i}^-|$. For this final layer of correction, we redistribute the remnant positive ECD values (i.e., $\rho_{B_j}^+ + \rho_{C_i}^-$) evenly across the voxels in $B_j$ and set the negative ECD values in $C_i$ to 0. This process is repeated for all clusters ($C$) until there are no negative ECD values remaining in the simulation cell (see Figure 2(b)).

## 3. FEATURE ENGINEERING & MODEL-BUILDING

### 3.1. COMPUTATION OF THE SPATIAL CORRELATIONS

The general mathematical framework for statistical quantification of a stochastic material structure $m(x, h)$ has already been established in prior work [29, 30, 33, 34, 53, 55, 56]. More specifically, the recently developed VAST framework [22, 38, 59] represents the molecular structure by its ECD field (i.e., $m(x, h) = \tilde{\rho}(x)$). Next, we featurize the pre-processed ECD field ($\tilde{\rho}_s$,) to capture the underlying translationally-invariant spatial patterns using the framework of 2-point spatial correlations, which will then serve as an input to our lean CNN model. Towards this goal, we compute the discrete autocorrelations of the discrete ECD field (specifically, its square root[1]) with Fast Fourier Transform (FFT) algorithms as

$$f_r = \frac{1}{|S|} \mathcal{F}^{-1}\left[\mathcal{F}\left(\sqrt{\tilde{\rho}_s}\right) * \mathcal{F}\left(\sqrt{\tilde{\rho}_s}\right)\right] \quad (2)$$

where $|S|$ is the total number of voxels in the simulation cell, $\mathcal{F}$ represents the Discrete Fourier Transform operation, $\mathcal{F}^{-1}$ represents the inverse Discrete Fourier Transform, and $*$ denotes the complex conjugate [26, 30]. The components of the autocorrelations vector defined by Eq. (2) form a high-dimensional feature vector, which can be used as an input to our proposed lean CNN models.

The discretized ECD fields utilized in this study were sourced from a curated repository of DFT simulation results, which employed different voxel lengths in the different simulations to accommodate the differences in the lattice parameters of the compounds studied. Our goal in the present study is to establish a CNN model that generalizes across diverse material classes (including different symmetry groups). As a result, we needed to standardize the input to the CNN model using a fixed voxel length for all compounds in our study. Note also that the voxel length can be standardized in the representation of the raw ECD field itself, i.e., before the computation of the 2-point spatial correlations as explored in prior work ([15], [22], [38]). In the present study, we opted to compute the 2-point correlations using the voxel lengths naturally implied in the ECD fields downloaded from the curated repositories, and subsequently express the required inputs to the CNN model (i.e., autocorrelations of the ECD field) using a standardized voxel length (i.e., kept the same for all compounds included in this study). This approach allows us to take full advantage of the periodicity of the ECD fields in the computation of the autocorrelations (using FFT algorithms). The interpolation of the computed periodic autocorrelations to the standardized uniform grid was accomplished using established trigonometric interpolation schemes [55, 56]. The standardized autocorrelations are denoted as $F_r$. Since the autocorrelations are typically much smoother compared to the original fields, our strategy to interpolate the autocorrelations as opposed to the raw ECD fields is expected to reduce significantly the interpolation errors involved.

In the present study, we adopted a standardized voxel width ($\lambda$) of 0.2 Å per voxel, and a uniform spatial grid of size $51 \times 51 \times 51$ for expressing the autocorrelations as the input to our lean CNN models. As a result, the spatial correlations included in this study covered vectors with components in the range of -5 Å to +5 Å. Although the compounds included in our study exhibited lattice parameters in the range from 2 Å – 12 Å, a large fraction of them had lattice parameters smaller than 5 Å. The decision to truncate the autocorrelations at ±5 Å was needed to limit the computational cost. The mathematical details of the trigonometric interpolation used in this work have been already presented in our prior work ([22]), Figure 3 provides an example of the data transformations performed on all the ECD fields downloaded and utilized in this work. This specific example corresponds to a SC unit cell of $Fe_3SnC$. For this compound, the ECD field from the public repository was available as an $80 \times 80 \times 80$ array covering a cubic volume with a side length of 3.83 Å. In the first step, the autocorrelations of the ECD field were computed using FFT algorithms ([27], [30], [34]), and expressed as an $80 \times 80 \times 80$ array covering vectors $r$ whose components lie within ±1.915 Å. In the second step, taking advantage of the periodicity of the autocorrelations and the trigonometric interpolation schemes, the autocorrelations are re-sampled as an array of size $51 \times 51 \times 51$ on a uniform grid covering vector components within ±5 Å. The value at the center of this array (corresponds to $r = 0$) reflects the average charge density for Fe$_3$SnC [26].

---

[1] Equation (1) has been defined with the square root of the ECD field $\tilde{\rho}_s$ so that the average charge density of the system is captured by $f_0$.

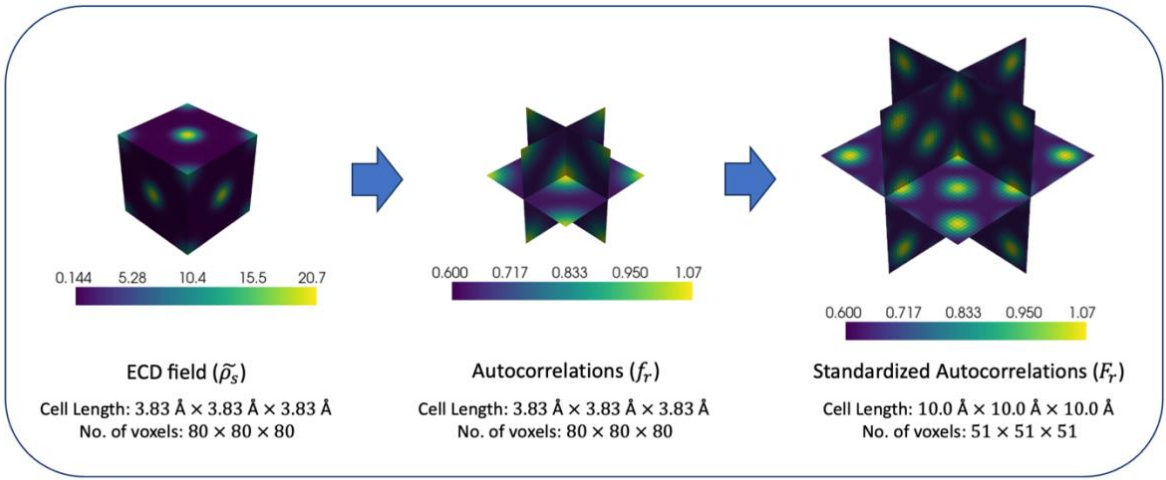

**Figure 3:** Illustration of the feature engineering steps performed on the ECD field of an example SC compound (Fe$_3$SnC). All the color bars represent eV/Å$^3$ values.

## 3.2. CONVOLUTIONAL NEURAL NETWORKS

Neural Networks (NNs) ([73], [74]) are powerful tools for learning complex nonlinear mappings between any selected set of inputs (i.e., features) and outputs (i.e., targets) encountered in solving engineering problems. A specific type, Convolutional Neural Networks (CNNs) ([36], [75], [76]), is ideally suited for problems where the inputs are images (i.e., spatial data). CNNs are typically designed by sequencing convolutional, pooling, and fully connected layers, which perform a variety of data transformations connecting the inputs to the outputs. The convolutional and fully connected layers introduce fittable parameters, which are calibrated to the training dataset. Some CNN architectures also employ strides with convolutions to achieve computationally efficient mappings between high-dimensional inputs and low-dimensional outputs (as is the case for the present application). Strides help coarsen inputs by reducing feature map dimensions during a convolution. The many options available in building CNNs result in several hyperparameters that affect model performance (usually quantified by higher accuracy or higher generalizability across the input domain). Examples of such hyperparameters include the number of layers, filter/kernels sizes, number of filters/kernels in each layer, choice of non-linear activation functions, and choice of loss functions. Tuning these hyperparameters is essential for producing a reliable CNN model. For example, while increasing a CNN's number of layers and the number of filters in each layer can improve learning, it also raises the risk of overfitting and increases computational costs. Therefore, it is important to attain the right balance in designing the CNN architecture for a given problem. CNNs have already been successfully applied to various materials science problems at the mesoscale ([35], [55], [57], [68], [77], [78], [79], [80], [81]) and the molecular scale ([32], [51], [70]). However, typical CNN applications often include fully connected (FC) layers, which dramatically increase the number of fittable parameters in the learned model. In this study, we address a homogenization problem that deals with 3D representations of the autocorrelations of the ECD field as the input. Since spatial variations in the input hold the keys to connecting to the target properties, it would make most physical sense to rely heavily on convolution layers and a final global average pooling (GAP) to predict the target properties of interest. Indeed, this simple strategy was successfully employed at the mesoscale in prior work [57]. Such simple CNN architectures comprising exclusively of only convolution layers and a final GAP layer produce very lean machine learning models (often less than 100K fittable parameters); these are referred as lean CNNs in this work. This reduction in model complexity is mainly attributed to using standardized autocorrelations ($F_r$) of the ECD field as input to the CNN models.

Because the mapping involved goes from a high-dimensional input (i.e., autocorrelations of ECD) to scalar outputs (i.e., bulk properties of interest), we have mainly focused on inverted architectures, where we gradually reduce the number of kernels in each subsequent layer. In study, we have explored a large number of possible architectures before settling on the best architecture for the present application. Models A, B & C (see Table 1) exemplify three of the architectures explored in this work. Models A & B represents single output CNNs that predict either the bulk modulus, or the total energy, as the individual scalar target. Model C represents a multi-output CNN that predicts both the bulk modulus and total energy simultaneously. One of our goals was to examine the relative performances of two independent single output CNNs versus a single multi-output CNN for our selected problem. This is interesting because the salient feature sets influencing the different outputs (i.e., bulk modulus and total energy) are expected to be dramatically different from each other. The comparison of the different CNN models produced in this work allowed us to examine critically the inherent feature-engineering capabilities of CNNs for diverse outputs in a single architecture. Furthermore, the CNN architectures examined in this work are deliberately simple, with a maximum of seven layers, and the best-performing model having six. Hence, we performed a manual search of architectures, given the limited complexity and the reduced computational cost compared to automated architecture searches. This approach allowed for effective intuition-driven optimization without the need for exhaustive automation, which we deemed unnecessary for a six-layer CNN with relatively fewer fittable parameters (see Table 1).

Table 1 describes the composition of the CNN architectures. Each layer is described using the notation p@b/c, which indicates a p × p × p (3-D) kernel, applied with a stride of c, and b channels. The default value of stride, when not mentioned, is one. For example, the notation 64@5/2 refers to 64 filters of size of 5 × 5 × 5, and a stride of 2. It should be noted that all the models we produced in this work utilize a significantly lesser number of fittable parameters (see the comparison with current benchmarks from literature in Table 1). Most interestingly, it will be seen that our lean Model A (see Figure 4), with the smallest number of fittable parameters, outperforms significantly some of the heavier models explored in this study.

For all the CNN models produced in this work, the target properties were normalized prior to model training and subsequently returned to their original scale after prediction. This was found to improve the convergence rate and ensured that both outputs received equal consideration in the loss function despite the large differences in their numerical values. The CNN models were produced using PyTorch [82]. PReLU (Parametric Rectified Linear Unit) [83] was adopted as the non-linear activation in this work, and is defined as

$$PReLU(x) = \begin{cases} x, & \text{if } x \geq 0 \\ \alpha x, & \text{otherwise} \end{cases} \quad (3)$$

where $\alpha$ is a fittable parameter associated with each filter (in addition to the weights and biases).

| Model | Zhao et al. [46] | Mirzaee et al. [54] | A | B | C |
|---|---|---|---|---|---|
| Input | $\rho_s$ | $\rho_s$ | $F_r$ | $F_r$ | $F_r$ |
| L1 | 5@32 | 3@16 | 5@64/2 | 5@128/2 | 5@64/2 |
| L2 | MaxPool | MaxPool | 3@32 | 3@64 | 3@32 |
| L3 | 2@64 | 3@32 | 3@16 | 3@32 | 3@16 |
| L4 | MaxPool | MaxPool | 3@8 | 3@16 | 3@8 |
| L5 | 3@64 | 3@64 | 1@1 | 3@8 | 1@2 |
| L6 | MaxPool | MaxPool | AvgPool | 1@1 | AvgPool |
| L7 | FC-1000 | FC-11200 | - | AvgPool | - |
| L8 | FC-128 | FC-2048 | - | - | - |
| L9 | FC-64 | FC-1024 | - | - | - |
| L10 | FC-32 | FC-8 | - | - | - |
| Outputs | 1 | 8 | 1 | 1 | 2 |
| Fittable Parameters | 2,360,736 | 2,129,672 | 80,825 | 310,201 | 80,834 |

**Table 1:** Examples of different CNN architectures explored in this work along with selected benchmarks from current literature.

Optimizing CNN performance required systematic hyperparameter tuning through multiple iterations, evaluating each configuration's impact on model fidelity for the specific task at hand. Beyond the hyper-parameters described earlier for the CNN architecture, we also tuned the initial learning rate and the number of epochs. The model's weights were optimized using stochastic gradient descent with the Adam Optimizer [84], and the learning rate was optimized using the Cosine Annealing Learning Rate Scheduler (with Warm Restarts) [85]. The initial learning rate was set to be $5 \times 10^{-3}$, and the training batch size was set to be 50. The Mean Squared Error (MSE) Loss function was used to train the models implemented in this work. Further, all CNN models were trained for several epochs until the respective loss functions converged to their minimum values. Through multiple trials, we established the number of training epochs at 650. By fixing the number of epochs, we enable a meaningful comparison of the performances of the various CNN architectures examined in this study. Dropout layers were included as a regularization technique, along with cross-validation measures ([86], [87]) to prevent overfitting and ensure model robustness.

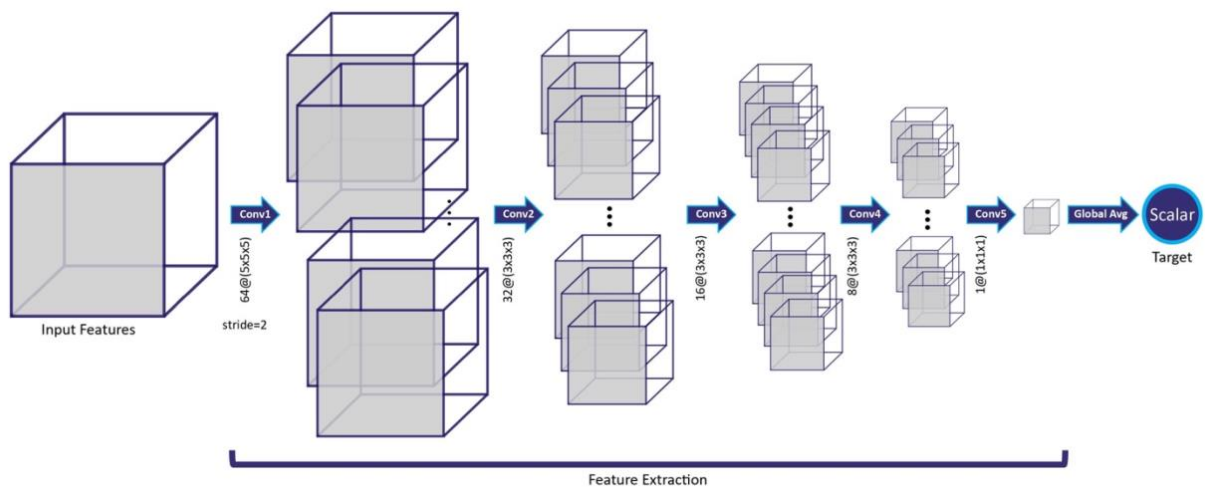

**Figure 4:** The CNN architecture (for Model A) implemented in this work.

## 4. RESULTS & DISCUSSION

The available limited dataset of 1410 datapoints exhibited high structural and chemical diversity, which necessitated careful consideration of data allocation into test, train, and cross-validation groups. To ensure the model captures this diversity and develops robust predictive capabilities, a large portion of the data was employed in training the models. In this work, we have utilized 90% of the data in training and cross-validation, and used only 10% for testing. This approach maximizes the use of available data for model development and ensures reliable performance assessment (through cross-validation).

The model building effort started with tuning of the CNN architecture using a 9-fold cross-validation (CV), where the 10% of the dataset was used in each fold for cross-validation. The CV errors were used as the main metric for the robustness of the trained model. In other words, model architectures producing lower CV errors were selected for further evaluation, and those exhibiting high CV errors were abandoned as over-fit models. Once the architecture was finalized, we used the full 90% training dataset for establishing the final version of the model. Model training was performed on a single NVIDIA A100 GPU with 32 GB of memory, and the training times averaged around 6-7 minutes for models A & C, and 11-12 minutes for Model B.

Table 2 summarizes the Mean Absolute Error (MAE) of the train and test sets for the lean CNN models produced in this work. For both bulk modulus and total energy predictions, Model A emerged as the best-performing model amongst all the CNN architectures investigated in this study (this includes the numerous other CNNs we produced beyond those reported in Table 2). The accuracy and robustness of model A for bulk modulus and total energy predictions can also be visualized in the parity plots shown in Figure 5. In this work, model performance was evaluated both in terms of accuracy (i.e., lowest MAE on the test set) and robustness (i.e., lowest difference between MAE values on the test and train sets). It can also be seen that Model B exhibits higher CV and test errors compared to Model A, even though it had almost four times as many fittable parameters. Clearly, Model B represents an overfit model.

| Model | A | | B | | C | |
|---|---|---|---|---|---|---|
| Properties | Bulk Modulus | Total Energy | Bulk Modulus | Total Energy | Bulk Modulus | Total Energy |
| Units | GPa | eV/atom | GPa | eV/atom | GPa | eV/atom |
| CV-MAE | 9.94 | 0.82 | 14.42 | 0.87 | 12.44 | 0.86 |
| Train-MAE | 7.49 | 0.73 | 14.57 | 0.81 | 10.74 | 0.74 |
| Test-MAE | 7.73 | 0.78 | 15.33 | 0.84 | 12.95 | 0.95 |

**Table 2:** Summary of the performance metrics for the lean CNN models trained in this work.

In order that we can compare the relative performance of the models for the two different targets selected for this study, we employed a Normalized MAE (NMAE) percentage defined as

$$NMAE = \frac{1}{N}\sum_{i=1}^{N}\left|\frac{S_i - \hat{S_i}}{S_{average}}\right| \times 100$$

where $S_i$ represents the ground-truth value for the target, $\hat{S_i}$ represents the corresponding model prediction, and $S_{average}$ denotes the ensemble average of the ground-truth values from a set of $N$ observations. Model A achieved a test NMAE of ~7% for bulk modulus predictions and a test NMAE of ~28% for total energy predictions. We observed that all the models produced in this work struggled to make accurate predictions for total energy. This is most likely because the total energy needs higher-order spatial correlations as inputs. Recall that we only utilized the 2-point spatial correlations (i.e., auto-correlations) as input to our CNN model. It is generally known that higher-order spatial correlations contain the critical predictors needed for the more complex mappings to the effective properties. In principle, CNNs can incorporate these higher-order correlations as additional channels in the input layer. It is clear from our investigation that CNNs are unable to automatically feature-engineer these higher-order spatial correlations from the auto-correlations we utilized as inputs to our models. This is particularly clear from the fact that a deeper CNN model, such as Model B, actually performs worse than the simpler Model A (see Table 2).

A comparison of the performance of Model C against the performance of Model A revealed that the multi-output strategy was ineffective for the targets selected in this study. We also observed that the performance of Model C did not improve even when we made the model bigger by adding more layers and channels. This implies that the multi-output CNNs were ineffective in learning the superset of features needed to predict both targets accurately. Indeed, the single-output CNNs were much more effective in learning the features separately for each target. The results of this study indicate clearly that a multi-output strategy is not likely to be advantageous in situations where the selected targets are expected to exhibit very different types of mappings with the inputs (i.e., S-P mappings). On the other hand, the multi-output strategy might still be beneficial for situations where the S-P mappings of interest have significant overlap.

The efficacy and simplicity of our model were evaluated by comparing its accuracy and complexity against conventional approaches, specifically random forest models utilizing "matminer" features based on chemical-formula-based descriptors [20]. Each sample in the "matminer" feature array comprised 65 scalar regressors, and the resulting models involved ~82K trainable parameters. Using the same dataset and targets as our lean CNN models, these overfit random-forest models achieved test-NMAEs of ~16% for bulk modulus and ~18% for total energy predictions, while the corresponding train errors were ~8% and ~9%, respectively. Additionally, the JARVIS leaderboard [88] demonstrates that larger datasets or simpler non-spatial features do not inherently ensure better predictions. Leveraging our effective featurization framework with a limited dataset, our model outperformed existing chemical-formula-based and graph-based surrogate models, achieving the best bulk modulus predictions to date (test-MAE: 7.73 GPa).

The computational savings achieved by the single-output lean CNN models (i.e., Model A) presented in this study are significant. Most impressively, Model A used <81K parameters to train, which represents a model with ~25-30 times lower number of trainable parameters used in similar models reported in recent literature ([46], [54]). Table 4 summarizes the main characteristics of Model A against these recent benchmarks. Despite being a leaner model, Model A significantly outperforms the benchmarks. This is impressive, keeping in mind that the benchmarks utilized larger datasets and covered only one crystal structure. Figure 6 provides histograms for the NMAE across the test set, for both the targets predicted by Model A. Most importantly, we note that ~55% of the test NMAEs lie below 10% for total energy predictions, while for bulk modulus predictions, ~70% of the test NMAEs lie under 10%. It is observed that ~5% of the bulk modulus predictions in the test set exhibit NMAEs above 30%, while for total energy predictions ~12% of the testing set has NMAEs above 30%. These high errors were caused by the fact that some of the chemical species encountered in the test set did not have adequate representation in the train set. Our analyses indicate that if a larger training dataset was available, the high test errors could be reduced significantly. We also observed no correlation between the high test errors and compounds with larger lattice parameters. This justifies our decision to limit the auto-correlations in a vector space with components ranging from -5 Å to +5 Å. This truncated set of auto-correlations appears to capture adequately the salient features needed for building the S-P models of interest in this work.

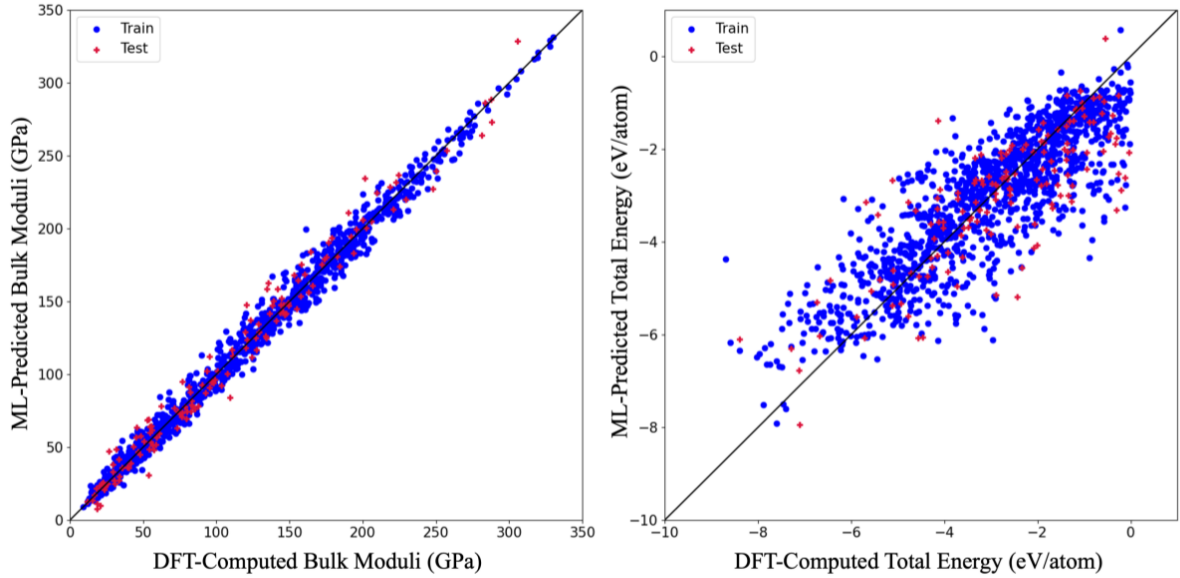

**Figure 5:** Parity plot of the ML-predicted (model A) v/s DFT-computed (a) Bulk Modulus and (b) Total Energy. The test points (red) are superimposed on the train points (blue).

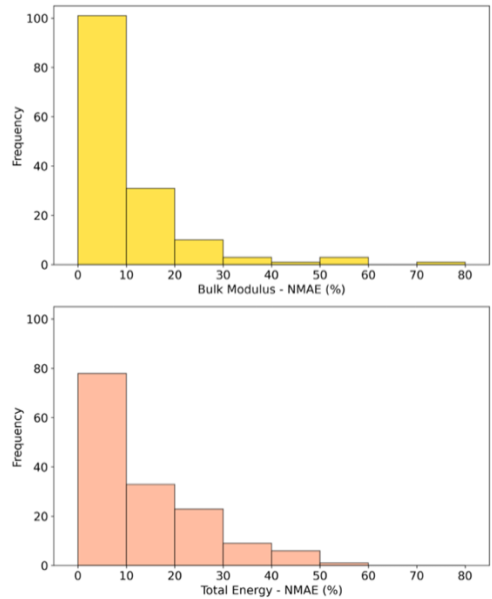

**Figure 6:** Normalized Mean Absolute Errors (NMAE) of both the target properties predicted by the best lean CNN model in this work (Model A).

| Model | Lean CNN (this work) | 3D-CNN (Zhao et al. [46]) | 3D-CNN (Mirzaee et al. [54]) |
|---|---|---|---|
| Input | $F_r$ | $\rho_s$ | $\rho_s$ |
| Total Dataset size | 1410 | 2170 | 2720 |

| Structural Complexities | FCC & SC | FCC | FCC |
|---|---|---|---|
| Trainable Parameters | 81K | 2.36M | 2.13M |
| RMSE (GPa) | 10.11 | 26.82 | 6.158 |

**Table 3:** Comparison of the lean CNN model trained in this work v/s benchmark model from literature ([46], [54]), for bulk modulus prediction.

## 5. CONCLUSIONS

This paper introduces a lean CNN architecture that significantly enhances the accurate and robust prediction of material properties (bulk modulus and total energy) from ECD fields. Our findings challenge the conventional wisdom that CNNs are primarily model-building techniques to circumvent explicit feature engineering. Instead, we demonstrate that the predictive capabilities of these networks can be substantially improved through physics-informed featurization, particularly with the use of spatial correlations. The lean CNNs described in this work are particularly useful in situations where there is limited data. The lean CNN's exceptional performance, achieving an NMAE of ~7% for bulk modulus predictions across 62 different elemental species, positions it as a new benchmark for future model-building efforts. This unprecedented accuracy, achieved using ECD fields as the sole structural descriptors of atomic space, opens new avenues for materials discovery and optimization. While the featurization framework introduced here has been demonstrated on cubic systems, it generalizes to arbitrary symmetries as it imposes no symmetry constraints, provided sufficient training data for diverse structures is available. Furthermore, the distribution of errors in predictions provides objective guidance on where additional training data should be included, as we observe that the target ranges are significant even with the data imbalance.

## 6. DECLARATION OF COMPETING INTEREST

The authors declare that no known competing financial interests have influenced the work reported in this paper.

## 7. ACKNOWLEDGEMENTS

PR and SK acknowledge support from NSF DMREF Award 2119640. The authors acknowledge Prathik Kaundinya and Conlain Kelly for their support and encouragement. This work used the Hive cluster, which is supported by the National Science Foundation under grant number 1828187. This research was supported in part through research cyberinfrastructure resources and services provided by the Partnership for an Advanced Computing Environment (PACE) at the Georgia Institute of Technology, Atlanta, Georgia, USA. K.C. thanks computational resources from the National Institute of Standards and Technology (NIST). This work was performed with funding from the CHIPS Metrology Program, part of CHIPS for America, National Institute of Standards and Technology, U.S. Department of Commerce. Certain commercial equipment, instruments, software, or materials are identified in this paper in order to specify the experimental procedure adequately. Such identifications are not intended to imply recommendation or endorsement by NIST, nor it is intended to imply that the materials or equipment identified are necessarily the best available for the purpose.

## 8. DATA AVAILABILITY

The dataset used for this study will be made available upon reasonable request, after internal review.